\documentclass[a4paper]{jpconf}
\usepackage{graphicx}
\def\bm#1{\hbox{\boldmath$#1$\unboldmath}}

\usepackage[fleqn]{amsmath}
\usepackage{amsmath}

\begin{document}

\title{Generalizing spin and pseudospin symmetries for relativistic spin 1/2 fermions
}

\author{P. Alberto$^1$, M. Malheiro$^2$, T. Frederico$^2$, A. de Castro$^3$}

\address{$^1$ Physics Department and CFisUC, University of Coimbra, P-3004-516 Coimbra,
Portugal}
\address{$^2$Instituto Tecnol\'ogico de Aeron\'autica, DCTA,
12228-900 S\~ao Jos\'e dos Campos, S\~ao Paulo, Brazil}
\address{$^3$Departamento de F{\'\i}sica e Qu{\'\i}mica,
Universidade Estadual Paulista, 12516-410 Guaratinguet\'a, S\~ao
Paulo, Brazil}

\ead{pedro.alberto@uc.pt}

\begin{abstract}
We propose a generalization of pseudospin and spin symmetries, the SU(2) symmetries of Dirac equation with scalar and vector mean-field potentials originally found independently in the 70's by Smith and Tassie, and Bell and Ruegg. As relativistic symmetries, they have been extensively researched and applied to several physical systems for the last 18 years. The main feature of these symmetries is the suppression of the spin-orbit coupling either in the upper or lower components of the Dirac spinor, thereby turning the respective second-order equations into Schr\"odinger-like equations, i.e, without a matrix structure. In this paper we use the original formalism of Bell and Ruegg to derive general requirements for the Lorentz structures of potentials in order to have these SU(2) symmetries in the Dirac equation, again allowing for the suppression of the matrix structure of the second-order equation of either the upper or lower components of the Dirac spinor. Furthermore, we derive equivalent conditions for spin and pseudospin symmetries with 2- and 1-dimensional potentials and list some possible candidates for 3, 2, and 1 dimensions. We suggest applications for physical systems in three and two dimensions, namely electrons in graphene.
\end{abstract}

\section{Introduction}
\vskip.3cm
Pseudospin symmetry has been a hot topic in nuclear physics since the late 60's,
when it was introduced to explain the near degeneracy of some single-particle
levels near the Fermi surface. Most of the formulations were then non-relativistic
but in 1997 Ginocchio was able to relate it
with a symmetry of the Dirac equation with scalar $S$ and vector $V$ mean-field
potentials such that $V=-S+ C$ where $C$ is a constant \cite{gino_1997}.
There is also a related symmetry, the spin
symmetry, that has been used explain the suppression of spin-orbit splittings
in states of mesons with a heavy and a light quark. Actually, both symmetries had been
described in the 70's independently by Smith and Tassie \cite{smith} and by Bell and Ruegg \cite{bell} as SU(2) symmetries
 of the Dirac equation with scalar (coupling with mass) and vector potentials (coupling with energy).
There several reviews of this subject, two extensive ones,  \cite{gino_rev_2005} and \cite{Liang_Meng_Zhou_rev},
and some more focused as \cite{conf_Praga} give a good overall
account of the many results produced in these last 18 years, especially in understanding the nature of the symmetries
in view of their application in several physical systems.

One remarkable feature of these symmetries is the suppression of either the spin-orbit or
the so-called pseudospin-orbit coupling that are present in the second-order equations
for the upper and lower Dirac spinor components, respectively. 
Since those terms arise from
the coupling of those spinor components in their first-order Dirac equations,
they have a non-trivial, i.e., different from identity,
matrix structure. Therefore, their suppression amounts to have the upper (spin symmetry) or lower
(pseudospin symmetry) spinors satisfying second-order equations of Shr\"odinger type, i.e, with no
 matrix structure. This means that one can suppress spin-orbit couplings even in a relativistic fermion system.
This may be surprising at first, since spin-orbit coupling is known to be
a relativistic correction to non-relativistic quantum mechanics with only vector potentials \cite{palberto_LS}, and actually
leads to conservation of orbital angular momentum in relativistic fermion systems with several kinds of potentials \cite{harm_osc_prc_2004_2006,spectra,pedro_pra}.

In this paper we show how to obtain other pairs of potentials in the Dirac equation that produce this same suppression,
under spin or pseudospin symmetry conditions. We extend this analysis to 2- and 1-dimensional potentials.
A detailed account of the procedure can be found in \cite{gen_sp_psd_pra}. We also suggest possible physical systems
to which the new symmetries may be applied to.

\section{General Spin and pseudospin symmetries in the Dirac equation}
\label{section 2}

\subsection{Generators of the spin symmetry}

The time--independent Dirac equation for a spin 1/2 particle with energy $E$, under
the action of an external hermitian $V$ potential (which may include a mass term) with a general Lorentz structure reads
\begin{equation}\label{eq:dirac_V}
H\psi=(\bm\alpha\cdot \hat{\bm p} + V)\psi=E\psi\ .
\end{equation}
where $\alpha_i$ ($i=1,2,3$) are the $4\times 4$ matrices related to the usual Dirac gamma matrices $\gamma^\mu$, $\mu=0,1,2,3$,  by $\alpha_i=\gamma^0\gamma^i$. Units $\hbar=c=1$ are used.

In order to find the requirements under which the potential $V$ meets the conditions for spin or pseudospin symmetries of
the Dirac equation (\ref{eq:dirac_V}), we follow closely the procedure of Bell and Ruegg \cite{bell}.

We consider operators $P_\pm=(I\pm O)/2$, with $I$ being the identity matrix
in spinor space, and a matrix is this space such that $O^2=I$. These have the projector properties
\begin{eqnarray}
\label{proj_prop1}
P_\pm P_\pm&=&P_\pm  \\
\label{proj_prop2}
P_\pm P_\mp&=&0 \ .
\end{eqnarray}
We also require that the anti-commutator $\{\alpha_i,O\}\ i=1,2,3$ is zero, such that
\begin{equation}\label{anticomm2}
P_\pm\alpha_i=\alpha_i P_\mp \quad i=1,2,3\ .
\end{equation}

In order that one may have spin or pseudospin-like symmetries, the potential $V$ has to have
the general form $V=V_O O +V_v I$, where $V_O$ and $V_v$ are functions of the coordinates,
such that it can be written as
\begin{equation}\label{V}
    V=V_+P_++V_-P_-\qquad{\rm with}\qquad V_\pm=V_v\pm V_O \ .
\end{equation}

We apply now the projectors $P_\pm$ to the Dirac equation (\ref{eq:dirac_V}) to get the two coupled equations
\begin{eqnarray}
\label{psi-}
 \bm\alpha\cdot\hat{\bm p}\,\psi_-+V_+\,\psi_+&=&E\psi_+ \\
\label{psi+}
 \bm\alpha\cdot\hat{\bm p}\,\psi_++V_-\,\psi_-&=&E\psi_- \ ,
\end{eqnarray}
where $\psi_\pm=P_\pm\psi$.

If we now set one of the potentials $V_\pm$, for example $V_-$, to a constant $C_-$, i.e., $V_v=V_O+C_-$,
we may apply $\bm\alpha\cdot\hat{\bm p}$ to equation (\ref{psi+}), and, by using the general property
$ \bm\alpha\cdot{\bm A}\,\bm\alpha\cdot{\bm B}={\bm A}\cdot{\bm B}+\rmi({\bm A}\times{\bm B})\cdot{\bm\Sigma}$
where $\bm\Sigma=\bm\alpha\times\bm\alpha/(2\rmi)=\gamma^5\bm\alpha$ is the spin matrix in four-dimensional spinor space,
we get
\begin{equation}\label{psi+2}
\hat{\bm p}^2\,\psi_+=(E-C_-)\bm\alpha\cdot\hat{\bm p}\,\psi_-=(E-C_-)(E-V_+)\,\psi_+
\end{equation}
using also eq. (\ref{psi-}).
This is a Schr\"odinger-type equation for $\psi_+$ with no matrix structure. Therefore, it is invariant under
the infinitesimal spin transformation
\begin{equation}\label{delta_psi+}
\psi_+\to \psi_++\delta\psi_+=\psi_++\frac{\bm\epsilon\cdot\bm\Sigma}{2\rmi}\psi_+
\end{equation}
The corresponding transformation for $\psi_-$ is
\begin{equation}
  \delta\psi_-= \frac{\bm\epsilon}{2\rmi}\cdot \frac{\bm\alpha\cdot\hat{\bm p}\,\bm\Sigma\,\bm\alpha\cdot\hat{\bm p}}
   {\hat{\bm p}^2}\,\psi_-\  .
\end{equation}
For the transformation of the full spinor $\psi=\psi_++\psi_-$, and defining
$\bm s={\bm\alpha\cdot\hat{\bm p}\,\bm\Sigma\,\bm\alpha\cdot\hat{\bm p}}\big/{\hat{\bm p}^2}$, we get
\begin{equation}\label{delta_psi}
    \delta\psi=\delta\psi_++\delta\psi_-=
    \frac{\bm\epsilon}{2\rmi}\cdot(\bm\Sigma\psi_++\bm s\psi_-)=
    \frac{\bm\epsilon}{2\rmi}\cdot(\bm\Sigma\, P_++\bm s\,P_-)\psi \ ,
\end{equation}
from which we can write the generator of this transformation as
\begin{equation}\label{gen_sym-}
\bm S_-=\bm\Sigma\, P_++\bm s\,P_- \ .
\end{equation}

One can obtain the second-order equation for $\psi_-$ from eqs. (\ref{psi-}) and (\ref{psi+}).
It reads
\begin{equation}\label{psi-2}
\hat{\bm p}^2\,\psi_-+\frac{1}{E-V_+}\big(\nabla V_+\times\hat{\bm p}\cdot{\bm\Sigma}-\rmi\,\nabla V_+\cdot\hat{\bm p}\big)\psi_-=(E-C_-)(E-V_+)\,\psi_- \ .
\end{equation}
If the potential $V_+$ is radial, in the second term in the left-hand side of the equation we can identify
 a spin-orbit coupling term and the Darwin term \cite{conf_Praga}.

 One can show that these generators satisfy a SU(2) algebra, i.e.,
 \begin{equation}\label{commut4}
  [(S_-)_i,(S_-)_j]=2\rmi\,\varepsilon_{ijk}(S_-)_k
   \end{equation}
and that they commute with the Hamiltonian $H_-=\bm\alpha\cdot \hat{\bm p} + V_+P_++C_-P_-$, provided  one has
$\{\bm\alpha,O\}=0$ or $[P_\pm,\bm\Sigma]=0$, which is actually a consequence of (\ref{anticomm2}) \cite{gen_sp_psd_pra}.

\subsection{Generators of the pseudospin symmetry}

Of course, we could as well have set instead $V_+$ in (\ref{V}) to a constant $C_+$. In that case, the roles of $\psi_\pm$  would
be reversed and one would have another symmetry whose generator would be
\begin{equation}\label{gen_sym+}
\bm S_+=\bm\Sigma\, P_-+\bm s\,P_+ \ ,
\end{equation}
which would commute with the Hamiltonian
\begin{equation}\label{H+}
    H_+=\bm\alpha\cdot\hat{\bm p}+V_-P_-+C_+P_+ \  .
\end{equation}
Similarly as the case for spin symmetry, one can show that these generators commute with $H_+$ and also satisfy a SU(2) algebra.

The second-order equations for the upper and lower spinors would then be
\begin{eqnarray*}
\hat{\bm p}^2\,\psi_+&+&\frac{1}{E-V_-}\big(\nabla V_-\times\hat{\bm p}\cdot{\bm\Sigma}-\rmi\,\nabla V_-\cdot\hat{\bm p}\big)\psi_+=(E-C_+)(E-V_-)\,\psi_+ \\
\hat{\bm p}^2\,\psi_-&=&(E-C_+)(E-V_-)\,\psi_- \ .
\end{eqnarray*}

This case,  $V_v=-V_O+C_+$, is usually known as the pseudospin symmetry case of the Dirac Hamiltonian, since we have a normal spin transformation in the lower component of the Dirac spinor which has an inverse parity relative to the upper component and to the whole spinor.

\section{Potentials allowing for general spin or pseudospin symmetries in 3, 2 and 1-dimensional space}

From the previous section, in order to have one of these two SU(2) symmetries, the matrix $O$ must satisfy the following relations:
\begin{enumerate}
  \item $O^2=I$
  \item \label{2}$\{\alpha_i,O\}=0 \qquad i=1,2,3 $.
\end{enumerate}
As explained before, the condition $[O,\Sigma_i]=0$ is also satisfied, as a consequence of condition (\ref{2}).
These requirements are satisfied by the Hermitian matrices $\beta=\gamma^0$ and $\rmi\gamma^0\gamma^5$.
The case of $O=\gamma^0$ leads to the well-known spin and pseudospin symmetries described in the Introduction.

If one weakens the second requirement, one can also consider the linear combination $O=\bm\lambda\cdot\bm\alpha$,
such that $\bm\lambda$ is a constant unit vector ($\bm\lambda\cdot \bm\lambda=1$). Indeed, if one requires that the anti-commutator
$\{\bm\alpha\cdot\hat{\bm p},O\}$ applied to $\psi$ is zero, then $\bm\lambda\cdot \hat{\bm p}\,\,\psi=0$. The effect, as before, is to
fulfill the property $P_\pm\bm\alpha\cdot \hat{\bm p}\,\,\psi=\bm\alpha\cdot \hat{\bm p}P_\mp\,\psi$. 
Then the third condition can be satisfied in a weak way, considering
transformations with infinitesimal parameters $\bm\epsilon$ such that
\begin{equation}\label{3req}
    [\bm\lambda\cdot\bm\alpha,\bm\epsilon\cdot\bm\Sigma]=2\rmi\,(\bm\lambda\times\bm\epsilon)\cdot\bm\alpha=0\ ,
\end{equation}
meaning that the vectors $\bm\lambda$ and $\bm\epsilon$ must be parallel. For instance, if $\bm\lambda=\hat e_z$,
i.e., $O\equiv\alpha_3=\gamma^0\gamma^3$, then one should have $\hat{p}_3\,\psi=0$ and $\bm\epsilon=\epsilon \hat e_z$. In this
case the symmetry generator would be the matrix $\Sigma_3$, generator of the two-dimensional rotation group in four-component
spinor space, which is a realization of the unitary group $U(1)$. Our problem would be 2-dimensional, i.e.,
the spinor (and potentials), would depend only on coordinates $x,\,y$.

Another possibility for $O$ would be the linear combination of the space components of the tensor operator in spinor space
$\gamma^0\sigma^{0i}=\rmi\beta\alpha_i$, i.e., $O=\rmi\beta\bm\lambda\cdot\bm\alpha$. The first requirement would be met
again by setting $\bm\lambda\cdot\bm\lambda=1$. The second requirements and the commutator condition $[O,\Sigma_i]=0$
would be met by setting, respectively,
\begin{flalign}\label{2req2}
  &\{\beta\bm\lambda\cdot\bm\alpha,\bm\alpha\cdot\hat{\bm p}\}\,\psi=
    \beta[\bm\lambda\cdot\bm\alpha,\bm\alpha\cdot\hat{\bm p}]\,\psi
    =2\rmi\,\beta\bm\lambda\times\hat{\bm p}\cdot\bm\Sigma\,\psi=0\ ,\\[2mm]
\label{3req2}
    &[\beta\bm\lambda\cdot\bm\alpha,\bm\epsilon\cdot\bm\Sigma]=\beta[\bm\lambda\cdot\bm\alpha,\bm\epsilon\cdot\bm\Sigma]
    =2\rmi\,\beta\bm\lambda\times\bm\epsilon\cdot\bm\alpha=0 \ .
\end{flalign}

The first condition would be satisfied if $\bm\lambda\times\hat{\bm p}\,\psi=0$ and the second one if $\bm\epsilon$ is
parallel to $\bm\lambda$. If one chooses again $\bm\lambda=\hat e_z$, this would give rise to a 1-dimensional potential,
depending only on $z$.

\section{Discussion and conclusions}

We have derived the general conditions under which a general potential plus a vector potential give rise to spin and pseudospin-like
symmetries in the Dirac equation, i.e., lead to a Schr\"odinger-like equation for the upper or lower component of the Dirac spinor.
In three-dimensional space, we showed that there are two potentials which satisfy those conditions: a scalar potential, giving rise to the usual the spin and pseudospin symmetries found independently by Smith and Tassie \cite{smith}, and  Bell and Ruegg \cite{bell} , and a pseudoscalar potential. In this last case the Dirac Hamiltonian would read
\begin{equation}\label{Hamilt_dirac_gamma5_V}
H=\bm\alpha\cdot \hat{\bm p} + \rmi\beta\gamma^5 V_{ps}+ V_v\ ,
\end{equation}
with $V_{ps}=\pm V_v\mp C_\mp$. In physical terms, this would correspond to a system of massless fermions interacting with
mean-field pseudoscalar and vector potentials which have the same magnitude up to a constant. One physical system in which this symmetry would be slightly broken would describe a fermion, say, a baryon, with a relatively small (effective) mass, interacting with a pion and $\omega$ meson.

On other hand, if we constrain the fermion motion to 2- and 1-dimensional space, there are additional
potentials for which these symmetries can be realized. The respective spin or pseudospin symmetric
Hamiltonians would look like
\begin{eqnarray}
\label{Hamilt_2}
  H_2&=&\alpha_x\hat p_x +\alpha_y\hat p_y+ \alpha_z V_z+ V_{2v} \\
\label{Hamilt_1}
  H_1&=&\alpha_z\hat p_z + \rmi\beta\alpha_z V_{t}+ V_{1v} \ ,
\end{eqnarray}
with 2- and 1- dimensional mean-field potentials such that $V_z(x,y)=\pm V_{2v}(x,y)$ and $V_t(z)=\pm V_{1v}(z)$. $V_z(x,y)$ is a (space) vector
potential, whereas $V_t(z)$ is a tensor potential.

Equation (\ref{Hamilt_2}) describes a 2-dimensional massless fermion system with a energy coupling potential and a vector ($\gamma^3$) potential.
 There is actually a physical system that this Hamiltonian can be applied to: electrons in graphene, the so called Dirac electrons. These effective particles can be described by a massless 3+1 Dirac equation within the framework of interacting quantum field theories (see e.g. \cite{JacPRL07,OliPRB11}). One example is the continuum spectrum of the Dirac electron interacting with two dimensional potentials embedded in a 3+1 space \cite{grafeno}. Again, in that theory one has in general also potentials with Lorentz structure other than
vector, and, in this case, the third component of a four-vector potential $(\gamma^0\gamma^3=\alpha_z)$
(note that the Lorentz character of $V_O O$ is given by its form in the covariant form of the Dirac equation, i.e., $\gamma^0 V_O O$).
 This opens the possibility of looking into what would be the effect of these symmetries on the continuum \cite{grafeno} and discrete spectrum of the Dirac electrons, as well as its breaking due the other potentials or to the fact that the condition $V_{2v}\pm V_z=C_\pm$ is not fulfilled. In this way, the Dirac electrons in graphene could be a tool to study the consequences of the  generalized spin and pseudospin symmetries in a controllable form. We leave for a future work such detailed investigation.

\ack
PA would like to thank the Universidade Estadual
Paulista, Guaratinguet\'a Campus, for supporting his stays in its
Physics and Chemistry Department. MM and TF would like to thank FAPESP for support under the thematic project 2015/26258-4 and CNPq for partial support.

\end{document}